\def\l{\lambda }  \def\r {\varrho }  \def\s{$\,$}  
\def\o{\omega }   \def\lc{\l _{{\rm crit}}}
\def\d{{\rm d}}   \def\e{\varepsilon }
\def\lsq{\vec l^{^{^{ \hspace{0.045in} 2}}}}
\def\psq{\vec p^{^{^{ \hspace{0.045in} 2}}}}
\font\tenyyy=cmcsc10 \def\yyy{\tenyyy}
\documentstyle[12pt]{article}
\newcommand{\simgeq}{\; \raisebox{-0.4ex}{\tiny$\stackrel
{{\textstyle>}}{\sim}$}\;}
\newcommand{\simleq}{\; \raisebox{-0.4ex}{\tiny$\stackrel
{{\textstyle<}}{\sim}$}\;}
\newcommand{\beq}{\begin{equation}}
\newcommand{\beqar}{\begin{eqnarray}}
\newcommand{\eeq}[1]{\label{#1} \end{equation}}
\newcommand{\eeqar}[1]{\label{#1} \end{eqnarray}}
\begin{document}
\hspace{5in} {\sc CNLS 94-02} \par \vspace{1in}
\baselineskip 30pt
\centerline{{\huge Shell Structures and Chaos in}}
\centerline{{\huge Nuclei and Large Metallic Clusters}}
\medskip
\baselineskip 20pt
\centerline{{\yyy
W.D.\s Heiss$^{\star}$, R.G.\s Nazmitdinov$^{\star \star}$
\footnote{on leave of absence from
Joint Institute for Nuclear Research,
Bogoliubov Laboratory of Theoretical Physics, 141980 Dubna, Russia}
and S.\s Radu$^{\star}$ }}
\medskip
\centerline{{\sl
$^{\star}$ Centre for Nonlinear Studies and Department of Physics}}
\centerline{{\sl
University of the Witwatersrand, PO Wits 2050, Johannesburg, South Africa }}
\centerline{{\sl
$^{\star \star}$ Departamento de Fisica Teorica C-XI }}
\centerline{{\sl Universidad Autonoma de Madrid, E-28049, Madrid, Spain}}
\begin{abstract}
A reflection-asymmetric deformed oscillator potential is analysed from
the classical and quantum mechanical point of view.
The connection between occurrence of shell structures and classical
periodic orbits is studied using the ''removal of resonances
method'' in a classical analysis. In this approximation, the { \it effective}
single particle potential becomes separable and the frequencies of
the classical trajectories are easily determined.
It turns out that the winding numbers calculated in this way are in good
agreement
with the ones found from the corresponding quantum mechanical spectrum
using the
particle number dependence of the fluctuating part of the total energy.
When the octupole term is switched on it is
found that prolate shapes are stable against chaos and can exhibit shells
whereas spherical and oblate cases become chaotic.
An attempt is made to explain this difference in
the quantum mechanical context by looking at the distribution
of exceptional points which results from the matrix structure of the
respective Hamiltonians. In a similar way we analyse
the modified Nilsson model and discuss its consequences for nuclei and
metallic clusters.
\end{abstract}
\textheight=22cm
\newpage
\voffset=-2.2cm
\section{Introduction}
The development of new experimental technique in recent years has
considerably
increased the accuracy of the measurement of nuclear properties at low
excitation energy. As a result, the superdeformed rotational bands have been
discovered \cite{Tw86} opening new perspectives on new phenomena like
the one of identical bands.  Low lying negative-parity
states, parity doublets, alternating parity bands with strong $E1$ -
transitions in radium and nearby nuclei \cite{G83} are another interesting
phenomena which could be considered as a manifestation of the octupole
deformation.
Possible occurrence of softness of superdeformed nuclei having the prolate
shapes with respect to the octupole deformation is one of the current topics
in nuclear structure physics (see for review \cite{Ab93}).

Ten years ago the pioneering experiments carried out by a group of
W.Knight \cite{Kn84} (see also \cite{He93,Bra93}) opened a new fascinating
field in study of many--body system, the atomic cluster phenomena,
an intermediate form of matter between molecules and bulk systems.
Experimental results on metallic clusters
concerning abundance spectra, ionization potentials, photoexcitation
etc was immediately interpreted in terms of electronic quantized motion in
a spherical effective potential.

Phenomena observed in many--body systems like nuclei and clusters could be
explained within the mean field approach based on the symmetry breaking
mechanism related to quantized single--particle motion. The quantisation
of a
system of Fermions moving in a common
potential leads to a bunching of levels in the single--particle spectrum,
known as shells.  The high level density  around the Fermi level (large
level bunching) corresponds to less stable system. When a spherical shell is
only partially filled, a breaking of spherical symmetry, resulting in an
energy gain, gives rise to a deformed equilibrium shape.
The basic concept of the deformation in the mean field approach is
the Jahn-Teller effect \cite{JT}, the mechanism first time proposed for
molecules, which leads to a spontaneous symmetry violation
(see discussion in \cite{NW94}).

On the other hand, the existence of magic numbers corresponding to
spherical and deformed nuclei/clusters could
be explained in terms of classical trajectories \cite{BB72} based on the
periodic orbit theory of Gutzwiller \cite{Gu71}.
According to the semiclassical theory \cite{BB72} the frequencies in
the level
density oscillations of single--particle spectra of nuclei are determined
by the corresponding periods of classical closed orbits. The short periodic
orbits give the major contribution to the gross shell structure
\cite{BM75,SM76}. Depending on the particular mean field potential a
deviation from spherical symmetry can lead to chaotic
motion in the corresponding classical problem, and the shell structure of the
corresponding quantum spectrum is affected or even destroyed depending on
the degree of chaos \cite{Gu90,Ar87,He94}.

Simple interpretation of the strong peaks, which are observed in the
abundance spectra of metallic clusters, in terms of the spectrum of
one-particle orbits in a spherical potential,
which are associated with periodic orbits of
the corresponding classical problem \cite{Ni90,Le93,PC93}, provide us
with the important key elements. The fact that the triangular orbits play
an important role for a family of quantum states with $\Delta l =3$ (here
$l$ is an orbital angular momentum, see \cite{BM75}) implies an importance of
octupole degree of freedom for configurations with partially filled shells
\cite{M94}.

In the present paper we consider a reflection-asymmetric axially deformed
oscillator potential and focus our analysis on the
distinction between orderly
and chaotic motion in many--body systems, like nuclei and metallic clusters.
One major result of the classical analysis
demonstrates that the prolate case including octupole deformation is
still quasi-integrable. In this way the quantum mechanical shell structure
found earlier \cite{He94} is
given a proper theoretical foundation \cite{HNR}.
Similarly, we analyse the modified Nilsson model including the $l^2$-term
and neglecting spin-orbit coupling, and the consequences
for nuclei and clusters. Quantum mechanical calculations
confirm that the $l^2$-term which gives rise to chaotic behaviour
in the corresponding classical problem \cite{HN94} may interfere with
the search of shell structure.

\section{The Model}

 We investigate the classical and quantum mechanical motion in
an axially symmetric deformed potential which
can be expanded as:
\begin{equation}
V( r , \theta)= \frac{m\omega_0^2}{2}r^2 \left( 1 + \alpha_2
P_2(\cos \theta)
+ \alpha_3 P_3(\cos\theta)+ ... \right).
\end{equation}
In this expansion the terms proportional to the components of the deformation
tensor $(\alpha_i)$ are the Legendre polynomials
of $i$-th order, respectively. In this paper we only consider
terms up to, and including, the third order. Rewriting the
potential in cylindrical
coordinates \cite{He94} we arrive at the form:
\begin{equation}
V(\r ,z) = \frac{m \omega^2}{2}\left( \r ^2 +
\frac{z^2}{b^2} + \lambda \frac{2z^3-3z\r ^2)}{\sqrt{\r ^2 + z^2}}
\right),
\end{equation}
where $\omega=\omega_0\sqrt{1-\alpha_2/2}$, $b=\sqrt{(1-\alpha_2/2)/(1+
\alpha_2)}$ and $\lambda=\alpha_3/(2-\alpha_2)$
We will call $b$ the quadrupole deformation and $\lambda$ the octupole
strength. For $b>1$ $(b<1)$ the potential is a deformed harmonic oscillator
of prolate (oblate) shape coupled to an additional octupole deformation
of strength $\lambda$. \par
Classically, the trajectories \cite{HNR} can be
calculated using the Hamilton formalism,
by integrating the following two equations of motion:
\[ \ddot{\r } = -\frac{\partial V(\r ,z)}{\partial \r } \]
\begin{equation}
\ddot{z}= -\frac{\partial V(\r ,z)}{\partial z}.
\end{equation}
Two aspects are important: First, because of
the axial symmetry, the $z$ component of the angular momentum $(l_z)$
is conserved.
As we will see in the next section, a finite value of
$l_z$ will not modify the winding numbers of the orbits. For this reason
we will only consider $l_z=0$. Second, we notice that the potential scales
as $V(\gamma\r ,\gamma z) = \gamma^2 V(\r ,z)$. This means that one
value of the energy will produce the same periods and trajectories
as any other energy value.

The use of the term $r^2P_3(\cos\theta)$ instead of
$r^3P_3(\cos\theta)$
in the potential ensures a proper
bound state problem for any value of the quadrupole deformation,
provided $|\lambda|$ is smaller than some $\lc$. Here, $
\lc$
is defined to be the least value of the octupole strength for which
the potential tends to -$\infty$ along one
direction in the $\r -z$ plane. Analytic expressions
for $\lc$ and the directions along which the potential opens its
valleys to $-\infty$ are given elsewhere \cite{HNR}.
There is a particular value of $b$, namely $b\approx 0.58$ for which valleys
along two directions $(\r =0)$ and $(\r \approx 0.4z)$ will open
simultaneously when $\lambda$ approaches its corresponding critical value.
We expect that this case will exhibit maximal chaotic behaviour.

\section{Classical Approach}

The coupling to the octupole potential renders the problem nonintegrable
\cite{He94}.
Despite this, it appears that in
prolate situations $(b>1)$, the quantum mechanical spectrum exhibits
shells \cite{He94}for some particular values of the octupole strength.
A periodic structure
which appears in the energy spectrum should be a reflection of one or more
shortest classical orbits which {\it dominate} in the phase space
\cite{BB72}.
 Here we present
a method which reveals that the classical
counterpart is quasi-integrable for prolate cases. It will also
establish that
the phase space structure is very close to that of
a potential without octupole term but with a larger value of the quadrupole
deformation $b$. \par
Our classical approach is based on  the ''removal of resonances'' method
developed in the secular perturbation theory \cite{LL81}. In the technique,
the Hamiltonian written in action-angle coordinates, is averaged over
the faster phase. Usually, prior to such
an operation, a canonical transformation is necessary in order to remove the
initial resonance from the unperturbed Hamiltonian (in our case the
Hamiltonian without the octupole coupling). In the new rotating frame, one
of the phases will only measure the slow variation of the variables about
the original resonance which now becomes a fixed elliptic point. The problem
is then treated by averaging over the remaining faster phase.
In the case of a super or hyper -deformed potential though, there already
appears to be a clear distinction between a slow and fast phase, therefore
the canonical transformation is unnecessary.
\par The complete
Hamilton function written in terms of the action-angle variables of the
unperturbed problem reads:

\begin{equation}
H(J_\r ,J_z,\theta_\r ,\theta_z)=
\omega\left(
J_\r  + \frac{1}{b}J_z +\lambda
\frac{\sqrt{bJ_z}\sin\theta_z ( 2 b J_z \sin^2 \theta_z -
3 J_\r  \sin^2 \theta_\r )}
{\sqrt{bJ_z\sin^2\theta_z + J_\r  \sin^2 \theta_\r }} \right)
\end{equation}
The frequencies of the two motions in $z$ and $\r $  can be expressed as:
\beqar
\omega_z(\vec J, \vec \theta)& =&
\frac{\partial H}{\partial J_z} \nonumber  \\
&=&\frac{\omega}{b} \left( 1 +
\lambda \frac{b \sin \theta_z (4b^2 J^2_z \sin^4 \theta_z +
6 b J_z J_\r  \sin^2\theta_\r  \sin^2 \theta_z -3 J^2_\r  \sin^2 \theta_
\r )}
{2\sqrt{bJ_z}(bJ_z\sin^2\theta_z + J_\r  \sin^2\theta_\r )^{\frac{3}{2}} }
\right)
\\
\omega_\r (\vec J, \vec \theta)& =&
\frac{\partial H}{\partial J_\r } \nonumber \\
&=&\omega\left( 1 - \lambda
\frac
{\sin\theta_z\sin^2\theta_\r \sqrt{bJ_z}
(8bJ_z\sin^2\theta_z + 3 J_\r  \sin^2 \theta_\r )}
{2\sqrt{bJ_z}(bJ_z\sin^2\theta_z + J_\r  \sin^2\theta_\r )^{\frac{3}{2}} }
\right)
\eeqar x

Obviously, the winding number of a trajectory defined as the ratio
$\omega_\r /\omega_z$ is essentially equal to $b$ when $\lambda$ is
small. This means that for $b$ sufficiently far from unity, there will always
be a fast and a slow moving phase, i.e. the averaging is performed over
the corresponding fast angle. \par

\subsection{Prolate Potential $(b>1)$ }

For prolate deformation the fast phase is $\theta_\r $. Averaging the
Hamilton function over $\theta_\r $, we are led to a new averaged
Hamiltonian $H_{av}$ which is independent of $\theta_\r $. This means
that the corresponding action $J_\r $ becomes a constant of motion   in
this approximation:
$J_\r  = J_\r (0)$. Introducing the notation
$\xi^2 = 2 J_\r (0)/(m\omega) = \r ^2(0) + p^2_{\r }(0)/(m\omega)^2$,
the averaged Hamiltonian reads in the original position momentum coordinates:
\begin{equation}
H_{av} = \frac{p^2_\r  + p^2_z}{2m} +
\frac{m \omega^2}{2} \left[
\r ^2 + \frac{z^2}{b^2} + \lambda \xi^2 \frac{{\rm sign}(z)}{2\pi}
\left( \frac{z^2}{\xi^2}{\rm K} ( -\frac{\xi^2}{z^2}) -
3\pi _2{\rm F}_1(\frac{1}{2},\frac{3}{2},2,-\frac{\xi^2}{z^2}) \right)
\right]
\end{equation}
Here, K and $_2{\rm F}_1$ are the complete elliptic function of the
first kind and the hypergeommetric function, respectively.
In this way, the problem is effectively reduced to two uncoupled
one dimensional cases.
A remainder of the actual coupling is the
fact that, through $\xi$, the $z$ motion still depends on the initial
conditions of the $\r $ motion.
The frequency $\omega_z$ is given by $\o _z=2 \pi/T$ where
\beq T=\sqrt{2m}\int_{z_{{\rm min}}}^{z_{{\rm max}}}{\d z
\over \sqrt{E_z-U(z)}} \eeq t
with $E_z=E-E_{\r }=E-m\o ^2\xi ^2/2$
and $U(z)$ is the {\it effective} potential of the $z$ motion and
is given by:
\begin{equation}
U(z) = \frac{m\omega^2}{2}
\left[ \frac{z^2}{b^2} + \lambda \xi^2 \frac{{\rm sign}(z)}{2\pi}
\left( \frac{z^2}{\xi^2}{\rm K} ( -\frac{\xi^2}{z^2}) -
3\pi _2{\rm F}_1(\frac{1}{2},\frac{3}{2},2,-\frac{\xi^2}{z^2}) \right)
\right]
\end{equation}
The winding numbers $\omega_\r /\omega_z$ can be evaluated numerically,
with $\omega_\r  = \omega$. It turns out that
for a given value of the octupole strength, the frequencies are virtually
independent of $\xi$,
and most of
the trajectories in the phase space will have basically the same winding
number. This number can be evaluated analytically at $\xi=0$:
\begin{equation}
\frac{\omega_\r }{\omega_z} =
\frac{b}{2}\left( \frac{1}{\sqrt{1+\lambda/\lc}} +
\frac{1}{\sqrt{1-\lambda/\lc}}\right) .
\end{equation}
We display in Fig.1 and Fig.2 the effective potential
and the variation of $\omega_\r /\omega_z$ for a few values of
$\xi$, respectively. As a function of $\lambda$ the winding number
increases monotonously. It means that for a prolate case, a
superposition of quadrupole and octupole deformation is practically
equivalent to a pure but larger quadrupole deformation.
Of particular interest is the existence of the short periodic orbits, as
they become important for the  quantum-classical correspondence.
For instance, in the superdeformed case $(b=2)$, according to formula (10),
the winding number becomes equal to $5:2$ at $\lambda = 0.66 \lc$ and should
have that value for most of the orbits. As a result, the quantum mechanical
spectrum is expected to exhibit a periodic structure \cite{He} of the
same kind as of
a pure quadrupole deformation with winding number $b=5:2$. Likewise,
when $\lambda=0.8\lc$, the ratio of the two frequencies becomes
$3:1$ and, as a consequence, a shell structure very similar to the case
$b=3$ and $\lambda=0$, is likely to occur. In Fig.3 we display the actual
phase space configuration $(z-p_z)$ for $\lambda = 0.66 \lc$ and
$\lambda = 0.8\lc$. Of importance is the large single separatrix
structure which covers most of the surface of section.
The corresponding periodic orbits lying in the centre of each separatrix
are displayed in Fig.4. Their winding numbers are $5:2$ and $3:1$,
respectively. After having tested numerically the winding number in different
regions of the phase space, it turns out that
with the exception of the innermost zones, the
winding numbers are very close to those given by formula (9) for about
$85 \%$ of the phase space.

The expression (10) for the ratio of the winding numbers is of course valid
for all values $b>1$ for which the approximation is meaningful. It turns
out that this is the case for $b\ge 1.5$. The significance of this statement
is far reaching: it means that even for irrational values of $b$,
where the unperturbed problem does not give rise to
closed orbits, the switching on of an appropriate octupole strength will
produce a situation of an unperturbed pure but larger quadrupole
deformation. For instance, $b=\sqrt {3}$ and $\l =0.56$ should essentially
yield the situation of pure superdeformation ($b=2$). The consequence for
the quantum spectrum will be discussed at the end of the following section.
\par As mentioned in the previous section, we restrict ourselves only to
zero value for the $z$ component of the angular momentum. A finite value
of $l_z$ does not change the result of the averaging procedure. This
implies that if at some value of $\lambda$ the quantum spectrum exhibits
a shell like structure for one value of the angular momentum, it
will exhibit the same type of structure for any angular momentum
and therefore produce genuine shells. We
return to this point in the next section.

\subsection{Spherical and Oblate Potential $(b\leq 1)$}

In these cases, the approximation procedure described above does not
achieve any decoupling
between $\r $ and $z$ motion. For an oblate potential, the faster phase
is $\theta_z$. Since the octupole part of the potential is an odd
function of $\theta_z$, the result of the averaging now vanishes.
If higher order terms in the Fourier expansion of
the octupole are considered,
the coupling
between the two degrees of freedom is still maintained; the attempt to
obtain an analytic
expression for the effective potential seems to stop here. The fact that
the problem remains coupled even perturbatively indicates
that chaotic behaviour may occur with increasing $\lambda$.
In the spherical case, one could in principle perform a canonical
transformation enforcing one of the two frequencies to become faster and
then averages over the faster phase.
However, the slower frequency turns out to be a complicated
function of time, in other words, the ratio $\omega_\r /\omega_z$ fails
to remain constant even for an individual orbit. Like in the oblate case,
the
spherical potential is expected to yield chaos for sufficiently
large values of
$\lambda$. For comparison, we display surfaces of sections in Fig.(5) for the
spherical and oblate case where the onset of chaos is obvious. We
mention here that the special case $b\approx 0.58$ requires the smallest
octupole strength to become chaotic. The onset of chaos is associated
with the disappearance of periodic (shell) structure in the corresponding
quantum mechanical spectrum. This feature will be dealt with in the next
section.

\section{Quantum Mechanical Approach}

We write the Hamilton operator in
the form $H_0+\l  H_1$ in a representation where $H_0$ is diagonal.
The basis chosen is referred  to as the basis
using the asymptotic quantum numbers $n_{\perp },n_z$ and $\Lambda $ where
$n_{\perp }=n_++n_-$ and $\Lambda =n_+-n_-$ \cite{BM75}. Here the quantum
numbers $n_+$ and $n_-$ are the eigenvalues of $(A_+)^{\dagger }A_+$ and
$(A_-)^{\dagger }A_-$,
respectively, where, in terms of the usual boson operators $a_x$ and $a_y$,
we use $A_{\pm}=(a_x\mp i a_y)/\sqrt{2}$. For a fixed value of
$\Lambda $
this leaves two quantum numbers (reflecting the two degrees of freedom)
to enumerate the rows and columns of the matrix problem. For $\Lambda =0$ the
diagonal entries of $H_0$ are thus $\e _{n_{\perp },n_z}^0=
\hbar \omega (n_{\perp } +1+(n_z+1/2)/b)$. The matrix elements
of $H_1$ are obtained from those of $z \sim (a^{\dagger }_z+a_z)$ and
$\r   ^2\sim (A_+(A_+)^{\dagger } +A_-(A_-)^{\dagger }+A_+A_-+(A_-)^{\dagger }
(A_+)^{\dagger })$. To get the matrix elements of $1/\sqrt{\r  ^2+z^2}$ in a
numerically consistent way, we first calculate the matrix elements $S_{m,n}$
of $\r  ^2+z^2$ from which the inverse square root is obtained using
$S^{-1/2}=U\cdot D^{-1/2}\cdot U^{\dagger }$ where $D=U^{\dagger }\cdot S
\cdot U$ is the diagonal form of the positive definite matrix $S$ and
$U$ is the orthogonal matrix which diagonalises $S$. To ensure also
numerically that $S$ has only positive eigenvalues it is important that
the matrix for $z^2$ is obtained by squaring $z$ and not by evaluating
analytically the matrix elements from $(a^{\dagger }_z+a_z)^2$;
inconsistencies are otherwise introduced due to truncation.
In this way, we also
ensure that the truncated matrices $S^{-1/2}$ and the representation of
$2z^3-3z\r  ^2$ commute.
\subsection{Energy Levels}
In Fig.(6a) we illustrate the spectrum so obtained as a function of $\l $ for
$b=2$. The shell structure at about $\l =0.63\lc $ and $\l =0.76\lc $
is clearly discernible. The fact that the structures do not occur exactly
at those values of $\lambda$ as calculated in section 3.1 is a consequence
of the approximation introduced by the ''removal of resonances'' method.
However, the agreement is very good, with an error of  only $5\%$.
While Fig.(6a)
presents the spectrum for $\Lambda =0$,
we illustrate in Fig.(6b) the whole spectrum which is a superposition of all
possible $\Lambda $-values. The shell structure is then more pronounced
which is expected since, according to the discussion in section 3.1, the
orbits will have the same winding numbers
independent of the angular momentum; as a consequence, the quantum spectra
will have shell structures similar to an unperturbed oscillator.
To contrast with the case $b=2$, we also display spectra for $b=1$ (spherical)
and $b=1/2$ (oblate superdeformed). As predicted by the classical analysis,
there are no obvious regions where periodic structures such as shells
would occur. The level statistics shows that such cases are
actually chaotic, in that their nearest neighbour distribution
approaches the Wigner surmise \cite{He94}. \par
\subsection{Total Energy Method}
To analyse the quantum spectrum we proceed in the orthodox fashion in that
the total energy $E_{{\rm tot}}(N,\l )=\sum _i^N \e _i(\l )$ is approximated
by a smooth average function $S(z,\l )$ and the fluctuating difference
\beq \delta E(N,\l )=E_{{\rm tot}}(N,\l )-S(N,\l ) \eeq  d
is then subjected to further investigation. The finding
of a suitable form for
the average function $S(z,\l )$ is facilitated in our case
as it is well known
that the leading term of $E_{{\rm tot}}(N,\l )$, as a function of
$N$, is proportional to $N^{4/3}$. We determine the five constants
$a_0(\l ),\ldots ,a_4(\l )$ in $S(z,\l )=\sum_{k=0}^4a_k(\l )z^{k/3}$ by a
least square fit which turns out to be perfectly satisfactory for all
values of $0\le \l <\lc $.

In Fig.(7a) the fluctuating part $\delta E(N,\l )$ is presented as a contour
plot and refers to $b=2$.
It displays the ranges $100\le N\le 700$ and $0\le \l <\lc $.
For $\l =0$ (the bottom horizontal line) we clearly discern the shell
structure of the plain deformed oscillator.
Note that the sharp minima (dark shadowing) occur at the positions $N$
where a shell is closed; the distances are proportional to $N^3$.
When $\l $ is switched on the shell structure persists to a great extent;
only when $\l $ approaches its critical value (top horizontal line) the
structure begins to be washed out. There are local minima discernible at
$\l /\lc \approx 0.76$. This is a reflection of the enhanced shell structure
discussed in the previous section.
For comparison we also display contour plots for $b=1.5$, $1$,
$0.58$ and $0.5$. As expected, the ordered formation of local minima
along a line $\lambda = const$ becomes lesser pronounced and eventually
disappears as the spherical and oblate region is approached.
The greatest extent of disorder is obtained for $b=0.58$ where the classical
case is maximally chaotic.
\par

\subsection{Magic Numbers}

The detailed shell structure is obtained
from the second derivative of $\delta E(N,\l )$ to which we turn next.
For $b=2$, we have plotted in Figs.(8)  the function
$g(E)=\delta E(N+1,\l )+\delta E(N-1,\l )-2\delta E(N,\l )$
versus $N^{1/3}$ for a few characteristic values of $\l $.
The peaks of the plots represent the magic numbers which characterise the
shells, and the heights of the peaks reflect the energy distance from one
shell to the next. In Fig.(8a) the essentially unperturbed result ($\l /\lc
=0.15$) is presented for demonstration. The second row displays
the results for the particular values of $\l $ for which the winding numbers
are 5:2 ($\l /\lc =0.63$) and 3:1 ($\l /\lc =0.76$), respectively. The magic
numbers and the heights of the peaks agree well with those which are obtained
from the unperturbed ($\l =0$) quadrupole deformed oscillators
(third row in Fig.(8)) with $b=5/2$ and
$b=3$, respectively, at least for $N\simleq 700,\, N^{1/3}\simleq 8.88$.
The agreement extends in particular to the respective occupation numbers,
i.e.\s the degeneracies; of course, the heights of the peaks do not show the
same regularity as the corresponding unperturbed problem; nevertheless,
even for the heights an overall agreement prevails when comparing with
Figs.(8e) and (8f) where the respective unperturbed structures are
displayed. For higher values
of $N$ we do get deviations which reflect upon the fact that the system is
nonintegrable and cannot give complete order in all its results. While the
agreement for lower values of $N$ was to be expected from the discussion in
the previous section, the extent of the agreement is rather remarkable,
especially for $\l \approx 0.76\lc $ where an astoundingly clean shell
structure reoccurs after it partially disappeared
for a somewhat smaller value
of $\l $. It is this recurrence of shell structure which gives
rise to the local minima in Fig.(7) as pointed out above. \par
One of the questions addressed in this paper is also the
possible occurrence of supershell effects when a quadrupole deformed harmonic
oscillator is perturbed by an anharmonic term, i.e.\s the octupole
deformation.
Of particular interest in this respect is Fig.(8b)
which refers to the intermediate value
$\l =0.7\lc $ where the genuinely different orbits with winding numbers
5:2 and 3:1 coexist. The long wave length fluctuation could well be
interpreted as a supershell structure, even though that it is not as clearly
pronounced as in a more transparent integrable case \cite{Ni90}. Yet the
difference to Fig.(8d) which refers to a larger value of $\l $ is striking.

\subsection{ Quantum-Classical Correspondence }

In Figs.(9) we display the square of the modulus of the Fourier transform
of the level density, i.e.\s the function
\beq F(t)=|\sum_n e^{i\e _n t}|^2. \eeq y
The spectrum is taken for $b=2$ at $\l =0.63\lc $ (Fig.(9a))
and at $\l =0.76\lc $
(Fig(9b)), both spectra refer to $\Lambda =0$ only. The pronounced peaks
can be directly associated with the periods of the classical 5:2 and 3:1
orbit, respectively, the periods obtained from Figs.(9) are in
perfect agreement with those of the corresponding classical orbits which are
found numerically by integrating Eqs.(3). This is a beautiful
demonstration of
Gutzwiller's trace formula \cite{Gu71}. As expected the frequencies
deviate considerably from the unperturbed values, i.e.\s from the frequency
associated with $b=2,\l =0$, but also from the frequencies
associated with $b=5/2$ or $b=3$. In units of the unperturbed value ($b=2$)
we find $T_{5:2}=1.2$ and $T_{3:1}=1.4$; the values are larger than unity in
accordance with Fig.(2). We mention that a similar situation is
encountered also in the
hyperdeformed prolate case. There the two main orbits have
the winding numbers $b=7:2$ (occurring at $\lambda = 0.6\lc$)
and $b=4:1$ (occurring at $\lambda=0.72\lc$), respectively. In a similar
vein, as was emphasised towards the end of Section 3.1, a shell structure
just like that of a pure superdeformed prolate case must occur for, say,
$b=\sqrt{3}$ and $\l \simleq 0.56$ \cite{Ar94}.
These findings make it quite clear that, as far as the
model is concerned, the spectrum alone cannot distinguish between an
additional octupole deformation and an unperturbed but larger quadrupole
deformation; additional experimental information is needed to settle this
point.

Again we stress that the high degree of order which prevails in the
superdeformed and hyperdeformed prolate case when the
octupole term is turned on, does not
exist in the corresponding oblate, in fact, not even in the spherical case.
There, chaotic behaviour becomes manifest for much lesser octupole strength,
which results in a complete disappearance of shell structure in the
quantum spectrum.

\section{Exceptional Points}

The spectra displayed in Figs.(6) exhibit avoided level crossings which
are related to the singularities \cite{He} of the energy in
the complex $\lambda$
plane. These singularities are known as exceptional points \cite{Kotze}.
They
represent the values of $\lambda$ where two energy levels coalesce when
continued into the complex plane.
The connection between the occurrence of avoided level crossings and
exceptional points is similar in nature to the connection between the
poles of a scattering function and the resonance structure of a cross
section. In the same way as the poles of the scattering function give
rise to the shape of the cross section, the exceptional points bring
about the shape of the spectrum. Their interplay alone \cite{Kotze} provides
the
mechanism for the signature of chaos in a quantum system. With the aid
of exceptional points,
criteria for quantum chaos can be found even when the classical counterpart
does not exist \cite{Kotze}.
\par
The levels are obtained by
solving the secular equation:
\begin{equation}
{\rm det} (E-H_0-\lambda H_1) = 0
\end{equation}
To enforce coalescence of the roots of Eq. (12) the additional algebraic
equation
\begin{equation}
\frac{d}{ d E} {\rm det}(E-H_0-\lambda H_1) = 0
\end{equation}
must be solved simultaneously.  Since the Hamilton operator is irreducible
with respect to symmetries, the fulfilment of the two equations
simultaneously is generically excluded for real finite $\lambda$, as this
would mean a genuine crossing for two levels. It has previously been
established that a high density of exceptional points is a prerequisite
for the occurrence of chaos in the energy spectrum \cite{Kotze}. Also,
it is known that
the statistical distribution of the real parts of exceptional points is
close to the distribution of avoided level crossings \cite{Kotze}.
It means that for
the particular values of $\lambda$ where the density of avoided level
crossings is high the energy spectrum generically obeys the same level
statistics as that ascribed to quantum chaos. We have analysed
the eigenvalue problem of the operator $H_0 + \lambda H_1$ beyond the
$\lc$ value. Although the region $\lambda > \lc$
is of no physical significance, it gives information about the
distribution of avoided level crossings.
To exclude genuine crossings between levels we display the $\Lambda=0$
subspace of the energy spectrum in Fig.(10). The spectrum is extended into
the physically forbidden region for the prolate
superdeformed $(b=2)$ potential.
It appears that the maximum of the distribution of level repulsions
occurs in the unphysical region for the prolate case in contrast to the
spherical and oblate case (see Figs.(6c,d)).
This can be understood since, unlike in the spherical and
the oblate situation, the first order perturbation term of the prolate
problem vanishes. As a result, the spectrum has a zero
derivative at $\lambda=0$, and the exceptional points cannot occur near
to the $\lambda=0$ axis, in fact most occur beyond $\lc$. In turn,
the occurrence of avoided level crossings in the physical region for
oblate and spherical potentials indicates the onset of chaos
for $\lambda<\lc$. While this was expected from the classical
behaviour, we see here the corresponding mechanism of the quantum mechanical
matrix problem at work. It is interesting to
note the connection between the symmetry breaking mechanism and the
distribution of level repulsions: the system aspires to break the spherical
symmetry where the density of avoided level crossings is high, so as to
keep order and stability.
The conclusion that the prolate case exhibits less chaos
for small values of $\lambda$ than the spherical and oblate case, can in
fact be drawn from the matrix structure alone, which in turn is directly
related to the gross distribution of exceptional points \cite{Kotze}.

\section{Modified Nilsson Model}

Used with a great deal of success in the description of deformed
nuclei, the modified Nilsson model has recently been extended
to metallic clusters \cite{Cl} (see for review \cite{He93,Bra93}).
In practical applications the Woods--Saxon potential, also used successfully
for the investigation of metallic clusters, has the disadvantage that it can
not be solved analytically, in contrast to the modified oscillator
potential \cite{Nil,Gus}. The level ordering in the Woods-Saxon case
falls between the soft-surface harmonic oscillator (HO) and the hard-surface
square well. The same level ordering is obtained in HO by the addition of
a term and reads in cylindrical coordinates
\begin{equation}
H = \frac{\psq}{2m} + \frac{m\omega^2}{2} \left(
\r ^2 + \frac{z^2}{b^2} \right) - v_{ll}  \hbar\omega \lsq
\end{equation}
Here, $\lsq$ is the square of the dimensionless angular momentum operator and
$v_{ll}$ is a constant. To avoid a general compression of the shells
produced by $\lsq$ alone, the average value of ${<\lsq>}_N$ is usually
subtracted. However, while in nuclear structure the
interest is focused on the lower lying levels in the single
particle potential,
the relevant energy range extends higher up for clusters. This
necessitates a re-discussion of the possible usefulness in
the application of this model to metallic clusters.
For the present discussion we leave out the
term ${<\lsq>}_N$ \cite{Brack2}. The classical
Hamilton function can then be defined:
\begin{equation}
H(\vec p, \vec r)= \frac{p_\r ^2 + p_z^2}{2m} + \frac{m\omega^2}{2}
\left(\r ^2 +\frac{z^2}{b^2} \right) - U\left( (\r  p_z - z p_\r )^2 +
l^2_z \right) + \frac{l^2_z}{\r ^2}
\end{equation}
Because of the axial symmetry, $l_z$ is conserved. Again,
the zero value of $l_z$ is sufficient for all relevant aspects of our
analysis. The classical problem is nonintegrable and gives rise to
chaotic motion \cite{HN94} if $U>0$ and $b\ne 1$.
We also note that the Hamilton function scales as:
$H(\alpha\vec p, \alpha\vec r) = \alpha^2 H(\vec p, \vec r)$, provided
the product $EU$ remains constant ($E$ is the energy).
Since the connection between $U$ and $v_{ll}$ is given by:
\begin{equation}
U = \frac{\omega}{\hbar} v_{ll},
\end{equation}
the classical behaviour for $EU=const$ will be reflected in the quantum
counterpart along the invariant lines $Ev_{ll} =const$.
To get an idea for which
energy and coupling strength signatures of chaos can be expected in the
quantum spectrum, a brief discussion of the classical phase space is
appropriate \cite{HN94}. In Fig.11 a section of the $z-p_z$ plane with the
$\r  =0$ plane is
displayed. The phase space is noncompact as can be seen from the figure
where the allowed regions are shaded.
The parameter $EU$ determines the structure of the phase space; if it
is larger than $\o ^2/4$ the two lines $z=\pm 1/\sqrt{2mU}$ intersect
the ellipse. In this case, chaos is clearly discernible inside the
ellipse for the classical motion even for small deviations from unity
of $b$.
If $EU<\o ^2/4 $ chaos occurs inside the ellipse only for $b\simgeq 1.5$. For
$E\le 0$ the ellipse is absent and the motion appears regular. Using the
expression (17) this translates for the quantum levels $E_N$
into $E_NU\simeq N\o ^2v_{ll}$ which means
that, for $v_{ll}=0.07$, chaotic behaviour is expected for $N\ge 3$
if $b=2$, while for, say, $b=1.05$ its onset should occur only as high as
$N\simgeq 20$.
Actual calculations fully confirm these expectations.
The fact that an arbitrary large negative energy is allowed
classically, is reflected also in the quantum spectrum; however, if the
matrix problem is truncated, there is a limitation on the eigenvalues of
$\lsq $; in any case, this part of the spectrum is virtually uncoupled to
positive energies and has no part in avoided level crossings.

\section{Summary}

  In the present paper we thoroughly investigated a simplified
model to analyse the nature of shell effects at
quadrupole/octupole axial deformations. We left out terms in the
Nilsson model, which, albeit physically important, are prone
to blur the analysis when the interest is focussed on the essentials,
in particular a distinction between orderly and chaotic motion
in many body systems, like nuclei and metallic clusters.
Contrary to the case of spherical
potentials  our Hamiltonian is nonintegrable. Using the method based on
the 'removal of resonances' approach \cite{LL81}, we established a
connection between shell structures in the quantum mechanical spectrum
and periodic orbits in the corresponding classical problem. In the prolate
case the classical problem can effectively be reduced to an integrable set of
equations, since the motion in the two coordinates becomes uncoupled.
This is in contrast to the spherical and oblate cases, where the motion
becomes chaotic when the octupole term is switched on. In the prolate case,
at particular values of the strength parameter $\l$ the octupole
term produces a motion
which resembles to a great extent that of a pure but more enhanced prolate
oscillator. A superposition of quadrupole and octupole deformations is
practically equivalent to a pure but larger quadrupole deformation.
This provides the classical explanation for the existence of
quantum shell structure for prolate/octupole deformed system within the model
\cite{He94,HNR}.
The fluctuating part of the energy has been extracted from the quantum
mechanical single particle spectrum using the procedure which is a variant
of the method discussed in \cite{BM75}. We have found
remarkable agreement between manifestations
of shell structure for the same values of $\l /\lc $ and the ones
which give rise to
large stability islands in the Poincar\'e surfaces
of sections relating to
the classical short orbits with winding numbers 2:1 ($\l \approx 0$), 5:2
$(\l /\lc = 0.63)$, 3:1 $(\l /\lc = 0.76)$.
In the intermediate case $(\l /\lc = 0.7)$ when the orbits
with winding numbers 5:2 and 3:1 coexist the long wave length fluctuation
could be interpreted as a supershell structure. But due to the narrowness of
these orbits the supershell structure is not as clearly pronounced as in a
more transparent integrable case \cite{Ni90}.
We placed our emphasis on the superdeformed case which is the
most interesting situation when the octupole term is switched on. Shell
structure is destroyed for smaller values of the octupole strength for
lesser quadrupole deformation or even more so for the spherical case.
Our results imply that the shell structure is supported by the superdeformed
prolate/octupole
in contrast to the oblate/octupole case. This conclusion was confirmed by
the analysis of exceptional points connected with
the occurrence of avoided level crossings in quantum spectra.
It appeared that
in the spherical and oblate case, the distribution of level repulsions,
which provide the mechanism for the signature of chaos in a quantum system,
occurs to a much greater extent in the physically relevant region
than in the prolate case.
The analysis of the modified Nilsson model without spin--orbit
coupling leads to the conclusion that
using the $\lsq $-term in models for metallic clusters with
the coupling strength similar to that in nuclear physics would
conflict, at least for larger deformations $b$, with the experimental
finding of shell structure. Only if the deformation $b$ is
very close to unity
can the
regular motion associated with shells prevail, provided the shell number
$N$ is not too large. In turn, only a much smaller coupling can yield
shell structure for larger deformations $b$. Since the $\lsq $-term has
only a phenomenological meaning, this finding simply calls for caution
when transferring models from nuclear to cluster physics.

\newpage
\centerline{\bf Figure captions} \par \vspace{0.5in}
{\bf Fig.1} The potential $U(z)$ in units of $m\omega^2/2$ for a few
values of $\lambda$  and an intermediate value of $\xi$. Variation of $\xi$
produces similar shapes. \par
{\bf Fig.2} The ratio $\omega_{\rho}/\omega_z$ as a function of $\lambda$ for
various values of $\xi$. A genuine $\xi$ dependence is discernible only
for $\xi >0.8 \xi_{\rm max}$. \par
{\bf Fig.3} Surfaces of section $(b=2)$ for $\lambda=0.63\lambda_{\rm crt}$
(top) and $\lambda=0.76\lambda_{\rm crt}$(bottom). \par
{\bf Fig.4} The major orbits with winding numbers $5:2$ $(\lambda/\lambda_{
\rm crt} = 0.63)$ and $3:1$ $(\lambda/\lambda_{\rm crt} =0.76)$. \par
{\bf Fig.5} Surfaces of section for $b=1$ (top) and $b=1/2$ (bottom) for
$\lambda=0.5\lambda_{\rm crt}$. \par
{\bf Fig.6} Spectrum $(b=2)$ for $\Lambda = 0$ (a) and for all $\Lambda$
(b) as a function of $\lambda$.
(c) and (d) are the $\Lambda=0$ spectra for $b=1$ and $b=1/2$,
respectively. \par
{\bf Fig.7} a.) Contours of $\delta E (N,\lambda)$ with $100<N<700$ as
absissa and $0<\lambda/\lambda_{\rm crt}<1$ as ordinate. Dark areas
represent minima. Volume conservation is taken into account in Figs.(7) and
(8). In b),c),d) and e) the same quantity is displayed for $b=1.5$,
$b=1$, $b=0.5$ and $b=0.58$, respectively. \par
{\bf Fig.8} Shell structure for fixed values of $\lambda$ and $b=2$. The
second derivative of $\delta E(N,\lambda)$ is illustrated versus $N^{1/3}$
for $\lambda /\lambda_{\rm crt} = 0$, $0.7$, $0.63$ and $0.76$ in (a), (b),
(c), and (d), respectively. In (e) and (f) the unperturbed $(\lambda=0)$
structure is displayed for $b=2.5$ and $b=3$, respectively. \par
{\bf Fig.9} Fourier transform of the modulus square of the level density
for $\Lambda = 0$, $b=2$ for $\lambda /\lambda_{\rm crt}=0.63$ (top)
and $\lambda /\lambda_{\rm crt}=0.76$ (bottom). \par
{\bf Fig.10} Spectrum $(b=2)$ for $\Lambda=0$ extended beyond $\lambda_
{\rm crt}=0.125$. Note that the distribution of the avoided level crossings
occurs largely for $\lambda>\lambda_{\rm crt}$. \par
{\bf Fig.11} Surfaces of section for the modified Nilsson model. Two
situations of accessible phase space (shaded) are displayed; left:
$|\lambda|< \omega^2/(4E)$, right: $|\lambda|> \omega^2/(4E)$

\end{document}